\newcommand{\Tr}{\mathop{\mathrm{Tr}}\nolimits}
\newcommand{\op}[1]{\hat{#1}}
\newcommand{\suma}[1]{\sum_{{#1}\in \mathbb{Z}}}
\newcommand{\integral}{\int_{2\pi}}

\documentclass[prl,twocolumn,showpacs,superscriptaddress]{revtex4}
\usepackage{amsfonts,amssymb,bm,graphicx,times,color}

\begin{document}

\title{Full quantum reconstruction of vortex states}

\author{I.~Rigas}
\affiliation{Departamento de \'Optica,
Facultad de F\'{\i}sica,
Universidad Complutense, 28040~Madrid, Spain}

\author{L.~L.~S\'{a}nchez-Soto}
\affiliation{Departamento de \'Optica,
Facultad de F\'{\i}sica,
Universidad Complutense, 28040~Madrid, Spain}

\author{A.~B.~Klimov}
\affiliation{Departamento de F\'{\i}sica,
Universidad de Guadalajara,
44420~Guadalajara, Jalisco, Mexico}

\author{J.~\v{R}eh\'{a}\v{c}ek}
\affiliation{Department of Optics,
Palacky University, 17. listopadu
50, 772 00 Olomouc, Czech Republic}

\author{Z.~Hradil}
\affiliation{Department of Optics,
Palacky University, 17. listopadu 50,
772 00 Olomouc, Czech Republic}
\date{\today}

\begin{abstract}
  We propose a complete tomographic reconstruction of vortex states
  carrying orbital angular momentum. The scheme determines the angular
  probability distribution of the state at different times under free
  evolution. To represent the quantum state we introduce a
  \textit{bona fide} Wigner function defined on the discrete cylinder,
  which is the natural phase space for the pair angle-angular momentum. 
  The feasibility of the proposal is addressed.
\end{abstract}

\pacs{03.65.Wj, 03.75.Lm, 42.50.Dv}

\maketitle

Phase singularities of wave fields were brought to attention in a
number of seminal papers~\cite{Dirac:1931,Nye:1974,Allen:1992}.
They manifest in electron wave packets~\cite{Bliokh:2007}, quantum
Hall fluids~\cite{Ezawa:2000}, supermedia~\cite{Salomaa:1987},
ferromagnets~\cite{Hubert:1998}, Bose-Einstein
condensates~\cite{Leggett:2001}, acoustical
waves~\cite{Hefner:1999}, and light fields~\cite{Allen:2003},
to cite some relevant examples. Special attention has been paid
to the particular case of vortices, which correspond to helicoidal
waves exhibiting a pure screw phase dislocation along the propagation
axis; i.e., an azimuthal phase dependence $e^{i \ell \phi}$. This means
that $\ell$ plays the role of a topological charge: the phase changes
its value in $\ell$ cycles of $2 \pi$ in any closed circuit about this
axis, while the amplitude is zero there.

One of the most interesting properties of vortices is that they carry
orbital angular momentum (OAM): indeed, the integer $\ell$ can be seen
as the eigenvalue of the OAM operator and its sign defines the
helicity or direction of rotation. In fact, OAM can be easily transferred, 
as demonstrated in a number of recent challenging experiments
with optically trapped microparticles~\cite{Simpson:1997}.

At the microscopic level, quantized vortices are now routinely
produced by, e.g., mechanical stirring of ultracold atomic clouds.
Remarkable images of large lattices containing hundreds of vortices 
in an Abrikosov-type triangular configuration have been
obtained~\cite{Abo:2001}. However, for a complete information about
the corresponding state, one needs to go beyond this mere photographic
picture and perform a full tomography.  Efficient methods of state
reconstruction are of the greatest relevance for quantum optics. 
Since the first theoretical proposals, this discipline has witnessed
significant growth~\cite{Paris:2004} and laboratory demonstrations of
state tomography are numerous and span a broad range of physical systems. 
The essence of vortices is their helicoidal structure, while other 
aspects, such as energy distribution, may be often ignored for many purposes: 
in this sense, a fully quantum reconstruction of the vortex content is 
pending.

Any reliable quantum tomographical scheme requires three key
ingredients~\cite{Hradil:2006}: the availability of a tomographically
complete measurement, a suitable representation of the quantum state,
and a robust algorithm to invert the experimental data. When these
conditions are not met, reconstruction becomes difficult, if not
impossible: this has been the case so far for vortex states. The
construction of a proper Wigner function for them (or any other
quasiprobability distribution in phase space) is still an open
question. Although some interesting attempts have been
published~\cite{Mukunda:1979,Bizarro:1994}, they seem of difficult
application for the problem at hand. The twofold goal of this paper
is precisely to fill this long overdue gap.  We will provide a simple
Wigner function with a clear geometrical meaning and also a universal
tomographic reconstruction scheme. One could think in using the
standard Wigner function for the transverse coordinate-momentum
variables instead. However, this is not the right answer in quantum
mechanics: such a representation conveys redundant information, it is
impossible to plot, and it hides all the relevant angular information.

We consider rotations by an angle $\phi$ generated by the OAM operator
along the $z$ axis, which for simplicity we shall denote henceforth as
$\op{L}$. We do not want to enter in a long and sterile discussion
about the possible existence of an angle operator. For our purposes
here the simplest choice is to use the complex exponential of the
angle $\op{E} = e^{- i \op{\phi}}$, which satisfies the commutation
relation $ [ \op{E}, \op{L} ] = \op{E}$.  The action of $\op{E}$ on
the OAM eigenstates is $\op{E} | \ell \rangle = | \ell - 1 \rangle$,
and it possesses then a simple implementation by means of a phase mask
removing a unit charge~\cite{Hradil:2006b}. Since the integer $ \ell $
runs from $ - \infty$ to $+ \infty$, $\op{E}$ is a unitary operator
whose normalized eigenvectors $|\phi \rangle$ describe states with
well-defined angular position. In the representation generated by
them, $\op{L}$ acts as $-i \partial_{\phi}$ (in units $\hbar =
1$). Note in passing that one could intuitively expect a Fourier-like
relationship between angle and OAM, which can be expressed in this
context as $ e^{- i \phi^\prime \hat L} | \phi \rangle = | \phi -
\phi^{\prime} \rangle$.

For the standard harmonic oscillator, the phase space is the plane
$\mathbb{R}^{2}$. In the case of angle and OAM the phase space is the
discrete cylinder $\mathbb{S}^{1} \times \mathbb{Z}$ (where
$\mathbb{S}^{1}$ is the unit circle).  While quasiprobability
distributions on the sphere are commonplace in quantum optics, to 
the best of our knowledge their counterparts on the cylinder have 
never been used to describe angular variables.  Given the key role 
played by the displacement operators in defining the Wigner function 
in the plane, we introduce a unitary displacement operator $\op{D} 
(\ell, \phi)$ on the discrete cylinder as
\begin{equation}
  \label{eq:Alg3}
  \op{D} (\ell, \phi)  = e^{i\alpha (\ell, \phi)} \,
  \op{E}{-\ell} e^{- i \phi \op{L})}  \, ,
\end{equation}
where $\alpha (\ell, \phi)$ is an undefined phase factor. Apart from
$2\pi$-periodicity in $\phi$, the requirement of unitarity imposes the
condition $  \alpha (\ell , \phi ) + \alpha (- \ell, - \phi )  =  - \ell \phi$.  
The displacement operators form a non-Hermitian orthogonal basis on 
the Hilbert space, in the sense that
\begin{equation}
  \label{eq:Alg5}
  \Tr [ \op{D} ( \ell , \phi ) \,
  \op{D}^\dagger (\ell^\prime, \phi^\prime ) ]
  =  2 \pi \delta_{\ell  \ell^\prime} \,
  \delta_{2\pi} ( \phi - \phi^\prime) \, ,
\end{equation}
where $\delta_{2\pi}$ represents the periodic delta function (or
Dirac comb) of period $2\pi$.

Next, we introduce the Wigner kernel as a kind of double Fourier 
transform of the displacement operators
\begin{equation}
  \label{eq:Wig1}
  \op{w} (\ell, \phi) =
  \frac{1}{(2\pi)^2}
  \suma{\ell^\prime} \integral d\phi^\prime
  \exp[-i ( \ell^\prime \phi - \ell \phi^\prime)]
  \op{D} (\ell^\prime, \phi^\prime) \, ,
\end{equation}
where the integral extends to the $2\pi$ interval within which the
angle is defined. The integral of $\op{w} (\ell, \phi)$ over the whole 
phase space yields unity and we may thus regard $\op{w} (\ell, \phi)$ 
as equivalent to the phase-point operators introduced by
Wootters~\cite{Wootters:1987}.  In addition, one can 
check that
\begin{equation}
  \op{w} (\ell, \phi) =
  \op{D} (\ell, \phi ) \, \op{w} (0,0) \,
  \op{D}^\dagger (\ell, \phi)    \, .
  \label{eq:Wig6a}
\end{equation}
Since for the plane and the sphere, the Wigner kernel can be 
seen as the transform of the parity by the displacement
operators~\cite{Klimov:2006}, this seems to call for interpreting
$\op{w} (0,0)$ as the parity over our phase space.

We next define the Wigner function of a quantum state described by 
the density matrix $\op{\varrho}$ as
\begin{equation}
  \label{eq:Wig7}
  W (\ell, \phi)  = \Tr [ \op{\varrho} \, \op{w} (\ell, \phi)  ] \, .
\end{equation}
Using the previous results for $\op{w} (\ell, \phi)$ one can show that
$ W (\ell, \phi)$ fulfills all the properties required for a reasonable
interpretation as a quasiprobability distribution.  Indeed, due to the
hermiticity of the Wigner kernel, $W (\ell, \phi )$ is real.  It also
provides the proper marginal distributions and it is covariant, which
means that if the state $\op{\varrho}^\prime$ is obtained from
$\op{\varrho}$ by a displacement in phase space $\op{\varrho}^\prime =
\op{D} (\ell^\prime, \phi^\prime ) \, \op{\varrho} \, \op{D}^\dagger
(\ell^\prime, \phi^\prime )$, then the Wigner function follows along
rigidly: $ W^\prime  (\ell, \phi) = W  (\ell - \ell^\prime , 
\phi - \phi^\prime )$. All these properties are fulfilled independently
of the choice of the phase $\alpha (\ell,\phi)$.

Since the displacement operators constitute a basis, we can write the 
expansion
\begin{equation}
  \label{eq:Wig9}
  \op{\varrho}  =  \suma{\ell} \integral d\phi \,
  \varrho (\ell, \phi) \op{D} (\ell, \phi )
\end{equation}
where $\varrho (\ell, \phi) = \Tr [ \op{\varrho} \, \op{D}^\dagger
(\ell , \phi) ]/(2\pi)$. In terms of $\varrho (\ell, \phi)$, the
Wigner function has the representation
\begin{equation}
  \label{eq:Wig11}
  W (\ell, \phi) = \frac{1}{2\pi}  \suma{\ell^\prime}
  \integral d\phi^\prime  \varrho (\ell^\prime, \phi^\prime) \,
  e^{i(\ell^\prime \phi  - \ell \phi^\prime)} \, .
\end{equation}

To work out explicit examples, one needs to fix once for all the
function $\alpha (\ell, \phi)$. One natural option is to set
$\alpha(\ell,\phi) = 0$ and then the Wigner kernel (\ref{eq:Wig1})
reduces to
 \begin{eqnarray}
    \label{eq:Xmpl4}
    \op{w} (\ell,\phi)  & = & \displaystyle
    \frac{1}{2\pi} \suma{\ell^\prime} e^{-2i \ell^\prime \phi}
    | \ell + \ell^\prime \rangle \langle \ell - \ell^\prime | \nonumber \\
    & +  & \displaystyle 
    \frac{1}{2\pi^2}  \suma{\ell^{\prime}, \ell^{\prime \prime}}
    \frac{(-1)^{\ell^{\prime \prime} - \ell}}{\ell^{\prime \prime} - \ell + 1/2} \nonumber \\
    & \times & 
    e^{- i (2 \ell^\prime + 1) \phi}
    |\ell^{\prime \prime} + \ell^\prime + 1 \rangle
    \langle \ell^{\prime \prime} - \ell^\prime | \, ,
  \end{eqnarray}
which agrees with the kernel derived by Pleba\'{n}ski and
coworkers \cite{Plebanski:2000} in the context of deformation
quantization on the cylinder.

For an OAM eigenstate $| \ell_0 \rangle$, we obtain
\begin{equation}
  \label{eq:ExAM1}
  W_{| \ell_0 \rangle} (\ell,\phi) = \frac{1}{2\pi} \delta_{\ell \ell_0} \, ,
\end{equation}
which is a quite reasonable Wigner function: it is flat in $\phi$ and
the integral over the whole phase space equals unity, reflecting the
normalization of $|\ell_0 \rangle$.

For an angle eigenstate $|\phi_0 \rangle$, we get
\begin{equation}
  \label{eq:ExAng2}
  W_{| \phi_0 \rangle} (\ell, \phi) = \frac{1}{2\pi} \,
  \delta_{2\pi}(\phi-\phi_0) \, .
\end{equation}
Now, it is flat in the conjugate variable $\ell$, and
thus, the integral over the whole phase space is not finite, which is
a consequence of the fact that the state $|\phi_0\rangle$ is
unnormalized.

The coherent states $| \ell_0, \phi_0 \rangle$ (parametrized by points
on the cylinder) introduced in Ref.~\cite{Kowalski:1996} (see also
Ref.~\cite{Kastrup:2006} for a detailed discussion of the properties
of these relevant states) satisfy
\begin{eqnarray}
  \label{eq:Xmpl5}
  \langle \ell | \ell_0 , \phi_0 \rangle & =  &
  \displaystyle
  \frac{1}{\sqrt{\vartheta_3 \left ( 0  \big | \frac{1}{e} \right )}}
  e^{-i \ell \phi_0} \, e^{-(\ell - \ell_0)^2/2} \, , \nonumber  \\
  & & \\
  \langle \phi| \ell_0 , \phi_0 \rangle & = &
  \displaystyle
  \frac{e^{i \ell_0 (\phi - \phi_0)}}
  {\sqrt{\vartheta_3 \left ( 0  \big | \frac{1}{e} \right )}}
  \vartheta_3\left(\frac{\phi-\theta}{2} \Big | \frac{1}{e^2} \right) \, ,
  \nonumber
\end{eqnarray}
where $\vartheta_3$ denotes the third Jacobi theta function.
One immediately finds that
\begin{equation}
  \label{eq:Coh5}
  \varrho (\ell, \phi)  =  
  \frac{e^{i \ell ( \phi/2 - \phi_0)}}{2\pi \vartheta_3
   \left ( 0  \big | \frac{1}{e} \right )} \,
  e^{- \ell^2/2 - i \ell_0 \phi} \, 
  \vartheta_3 \left( \frac{\phi}{2} + \frac{i\ell}{2}\Big|
    \frac{1}{e} \right) \, ,
\end{equation}
which gives
\begin{equation}
  \label{eq:1}
  W_{|\ell_0 , \phi_0 \rangle} (\ell,\phi) =
  W^{(+)}_{|\ell_0 , \phi_0 \rangle} (\ell,\phi) +
  W^{(-)}_{|\ell_0 , \phi_0 \rangle} (\ell,\phi) \, ,
\end{equation}
with
\begin{eqnarray}
  \label{eq:Coh1}
  W^{(+)}_{|\ell_0,\phi_0 \rangle} (\ell,\phi) &  =  &
  \frac{1}{2\pi\vartheta_3 \left ( 0  \big | \frac{1}{e} \right )} \,
  e^{-(\ell-\ell_0)^2}
  \vartheta_3\left(\phi-\phi_0 \Big|\frac{1}{e}\right) \, , \nonumber \\
  \\
  W^{(-)}_{| \ell_0 , \phi_0 \rangle} (\ell,\phi) & = &
  \frac{e^{i(\phi-\phi_0)-1/2}}{2\pi^2 \vartheta_3
    \left ( 0  \big | \frac{1}{e} \right )}
  \vartheta_3\left( \phi-\phi_0 +i/2 \Big|\frac{1}{e} \right) \nonumber \\
  \displaystyle
  & \times &
  \suma{\ell^\prime} \frac{(-1)^{\ell^\prime - \ell + \ell_0} \,
    e^{-\ell^\prime {}^2} - \ell^\prime}{\ell^\prime + \ell_0 - \ell + 1/2}  \, .
  \nonumber
\end{eqnarray}
In spite of the fact that these states may be regarded as a basic
set to construct the Wigner kernel, their Wigner function has no
simple closed form, but it splits into two different functions with
periods $\pi$ and $2\pi$, respectively.

\begin{figure}[t]
\includegraphics[width=0.80\columnwidth]{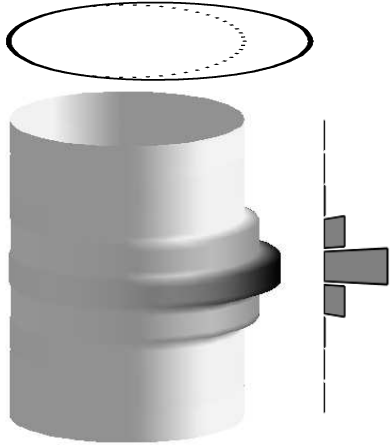}
\caption{Plot of the Wigner function for a coherent state with
$\ell_0 =0$ and $\phi_0 = 0$. The cylinder extends vertically
from $\ell = -4 $ to $\ell = +4$. We show the two corresponding
marginal distributions.}
\label{fig:mathieu}
\end{figure}

In Fig.~1 we show the Wigner function for the coherent state
$| \ell_0 = 0, \phi_0 = 0\rangle$ plotted on the discrete cylinder. 
We can see a pronounced peak at $\phi=0$ for $\ell=0$ and slightly smaller 
ones for $\ell=\pm 1$. A closer look at the picture reveals also a remarkable 
fact: for values close to $\phi=\pm \pi$ and $\ell= \pm 1$, the Wigner function
takes negative values.  Actually, a numeric analysis suggests the
existence of negativities close to $\phi = \pm \pi$ for odd values of
$\ell$. We also plot the marginals obtained from Eq.~(\ref{eq:1}) 
by integrating over $\phi$ or summing over $\ell$.

As our last example, we look at the superposition states
\begin{equation}
  \label{super}
  |\Psi \rangle  = \frac{1}{\sqrt{2}} ( | \ell_0  \rangle +
  e^{i \phi_0} \,  | - \ell_0  \rangle) \, , 
\end{equation}
which have been proposed for applications in quantum
experiments~\cite{Vaziri:2002}. The analysis can be carried out for
more general superpositions, but (\ref{super}) is enough to display
the relevant features. The final result is
\begin{equation}
  \label{Wsuper}
  W_{| \Psi \rangle} (\ell,\phi)   =   \frac{1}{4 \pi} ( \delta_{\ell \ell_0} 
  +  \delta_{\ell \, - \ell_0} ) +  \frac{1}{2 \pi}  
  \cos (\phi_0 - 8 \phi) \, \delta_{\ell 0} \,  .
\end{equation}
The Wigner function presents then three contributions: two flat slices
coming from the states $ | \ell_0 \rangle $ and $ | - \ell_0 \rangle$
and their interference located at the origin (see Fig. 2).

\begin{figure}[t]
  \includegraphics[width=0.75\columnwidth]{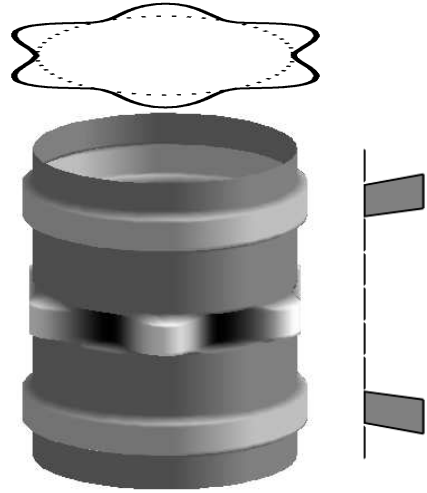}
  \caption{Plot of the Wigner function (and the corresponding
    marginals) for a superposition state~(\ref{super}) with $\ell_0
    =3$ and $\phi_0 = \pi$. and its marginal distributions. The
    cylinder extends vertically from $\ell = -4 $ to $\ell = +4$.}
\end{figure}

To complete our theory we also propose a reconstruction scheme for
these sates. A reconstruction of $\op{\varrho}$ (or, equivalently, of
its Wigner function) is tantamount to finding the coefficients
$\varrho (\ell,\phi)$. To this end, we need a tomographical
measurement that allows us to reconstruct $\varrho (\ell,\phi)$.  
For $\ell = 0$, the coefficients $\varrho (0, \phi) $ read as
\begin{equation}
  \label{eq:Tom1}
  \varrho (0, \phi) = \frac{1}{2\pi}  \suma{\ell}
  \langle \ell | \op{\varrho} | \ell \rangle \, e^{i \ell \phi} \, ,
\end{equation}
where we have made use of the fact that the undetermined phase
$e^{-i\alpha (0,\phi)}$ can be set to 1 for all values of $\phi$,
for it corresponds to displacements along one of the coordinate
axes and no additional phase should be acquired.   

For $\ell \neq 0$, we introduce the operator $\op{U}_{t} = \exp(-i t
\op{L}^2 /2)$.  If we recall that the generic Hamiltonian of a quantum
rotator is $\op{H} = \op{L}^{2}/2I$ ($I$ being the rotational moment of 
inertia), the operator $\op{U}_t$ is
simply the free propagator of the system (in appropriate units). This
quantum rotator Hamiltonian describes a variety of situations,
such as molecular rotations~\cite{Mouritzena:2006}, single-photon
OAM~\cite{Molina:2004} or the azimuthal evolution of
optical beams~\cite{Rehacek:2008}. For all of them, our proposal for
$\omega(\phi, t)$ is precisely to measure the angular distribution 
after a free evolution $t$, that is,
\begin{equation}
  \label{eq:Tom7}
  \omega( \phi, t)  =
  \langle\phi(t) | \op{\varrho} |\phi(t) \rangle =
  \langle \phi |\op{U}_{t}^\dagger \,  \op{\varrho} \,
  \op{U}_{t}  | \phi \rangle \, .
\end{equation}
In other cases, the scheme also works appropriately provided
one can experimentally implement the action of $\op{L}^2$. 
Using the representations (\ref{eq:Wig9}) for $\op{\varrho}$ and
(\ref{eq:Alg3}) for $\op{D}$, $\varrho(\ell,\phi)$ turns out to be
\begin{equation}
  \label{eq:Tom14}
  \varrho (\ell,\phi) = \frac{1}{2 \pi} e^{-i\alpha(\ell,\phi)}
  \integral d\phi^\prime \,  e^{-i\ell \phi^\prime} \,
  \omega (\phi^\prime,\phi/\ell ) \, .
\end{equation}
Plugging these coefficients into Eq.~(\ref{eq:Wig11}), we get the
reconstruction of the Wigner function as
\begin{eqnarray}
  \label{eq:Tom15}
  W(\ell,\phi) & = &  \displaystyle
  \frac{1}{2\pi}
  \langle \ell| \op{\varrho} | \ell \rangle
  +  \frac{1}{(2\pi)^2}
  \sum_{\ell^\prime\in\mathbb{Z}\atop \ell^\prime \neq 0}
  \integral d\phi^\prime \, d\phi^{\prime \prime}
  e^{-i\alpha (\ell^\prime,\phi^\prime)} \,
  \nonumber \\
  & \times &
  e^{i[\ell \phi^\prime - \ell^\prime (\phi - \phi^{\prime \prime})]} \,
  \omega (\phi^{\prime \prime}, \phi^\prime/\ell^\prime ) \, .
  \label{eq:Tom16}
\end{eqnarray}
Consistently, the reconstruction procedure itself does not depend
on the undetermined phase $\alpha(\ell,\phi)$ of the displacement
operator, while the Wigner function does have such a dependence.
The phase factor acts as a metric coefficient for the mapping
from Hilbert space onto the phase space.  We recall that for the
harmonic oscillator, one can recover the Wigner function via an
inverse Radon transform from the quadrature probability
distribution~\cite{Vogel:1989}. Equation~(\ref{eq:Tom15}) is
then the analogous  for our system.

As a rather simple yet illustrative example, let us note that for 
the vortex state $| \ell_0 \rangle$, $\op{U}_t$ is diagonal, so the
tomograms $\omega (\phi^\prime, \phi^\prime/\ell)$ are independent of
$\phi$ and $\ell$ and all of them equal to $1/(2\pi)$.  Performing 
the integration in Eq.~(\ref{eq:Tom16}) we obtain precisely the 
Wigner function (\ref{eq:ExAM1}).

Finally, for the feasibility of the proposed scheme, we need the
projection onto the eigenstates $| \phi \rangle$.  Since these 
states correspond to the measurement of a continuous variable, 
such a measurement can be done only approximately. A good 
approximation  seems to be the projection onto the wedge 
states, although other experimental schemes  are also 
available~\cite{Dooley:2003}.  According  to Eq.~(\ref{eq:Tom7}), 
this projection, together with the  determination of $t$, are 
the two main sources of measurement  error. Afterwards, the 
reconstruction errors for each particular  experimental 
implementation should be evaluated following the 
standard recipes developed in, e.g., Ref.~\cite{Rehacek:2008b}.

In summary,  we have carried out a full program for the
reconstruction of generic vortex states, including a complete
phase-space description in terms a \textit{bona fide} Wigner
function. Though the implementation of this scheme
may differ depending on the system under consideration, our
formulation provides a common theoretical framework on the
Hilbert space generated by the action of angle and angular
momentum. Experimental demonstrations of the method
in terms of optical beams are presently underway in our
laboratory and will be reported elsewhere.

\acknowledgments

We acknowledge discussions with Hubert de Guise, Jose Gracia-Bondia,
Hans Kastrup, Jakub Rembielinski and Krzysztof Kowalski.  This work
was supported by the Czech Ministry of Education, Project
MSM6198959213, the Czech Gran Agency, Grant 202/06/0307, the 
Spanish Research Directorate, Grant FIS2005-06714 and 
the Mexican CONACYT, Grant 45705.


\end{document}